# Design Principles and Observable Indicators for AI-Enabled Pedagogical Accompaniment: Evidence from the Amico Dual-Mode Prototype in Italy and China

**Pier Paolo Benedetti**
Doctoral Program in Education
Universidad de Almería, Almería, Spain
Zhejiang Yizhong Intelligent Technology Co., Ltd., Hangzhou, China
p.benedetti@edulife.it

**Status**
Accepted and presented at the 2026 International Conference on Artificial Intelligence and Education (ICAIE 2026). Proceedings forthcoming.

# Design Principles and Observable Indicators for AI-Enabled Pedagogical Accompaniment: Evidence from the Amico Dual-Mode Prototype in Italy and China

Pier Paolo Benedetti
*Doctoral Program in Education*
*Universidad de Almería,* Almería, Spain
*Zhejiang Yizhong Intelligent Technology Co., Ltd., Hangzhou, China*
p.benedetti@edulife.it

***Abstract*—AI-enabled systems are increasingly introduced into educational contexts, yet their effectiveness depends less on technological sophistication than on the quality of pedagogical mediation, ethical constraints, and context-sensitive design. This paper proposes a replicable framework for AI-enabled pedagogical accompaniment, grounded in a human-in-command approach in which adult responsibility remains central and AI functions as an enabling, non-substitutive infrastructure. Building on the Amico project, we operationalize the concept of a relational bridge as a sequence of micro-mediations that lower the threshold of access to educational relationships and facilitate transitions toward meaningful human interaction with teachers, peers, and communities of practice. The contribution synthesizes a set of design principles, including transparency of system identity and limits, scaffolding toward human contact, maieutic questioning, prevention of dependency dynamics, and data minimization, and maps them to observable indicators suitable for real educational settings. The paper also outlines an initial cross-context exploration of the prototype in Italy and China and discusses how the two interaction modes, AmicoMio (structured, task-oriented) and AmicoTuo (reflective, supportive), can be used as complementary pedagogical mediations. Pilot observations and participant feedback suggested feasibility and perceived usefulness in vocational contexts, motivating the present framework, informing the subsequent doctoral research program, and supporting the proposed collaborative research agenda.**

## I. Introduction

In recent years, artificial intelligence has been increasingly integrated into educational contexts, with growing expectations regarding personalization, efficiency, and scalability of learning processes. AI-enabled systems are now employed to support tutoring, assessment, feedback, and learning analytics across a wide range of educational settings [1] including emerging agent-based and large language model–driven educational systems [2]. However, alongside these developments, converging signs of relational fragility have emerged: difficulties in dialogue between adults and young people, reduced communicative initiative, fragmentation of learning motivation, and a growing sense of meaning fatigue among students. These phenomena are observable across different cultural and institutional contexts, albeit with varying interpretations and responses [3].

In both Western and East Asian educational systems, purely technological or performance-driven approaches have shown limited capacity to address such challenges [3], [4]. While AI can enhance access to information and automate certain instructional functions, it does not, by itself, regenerate the educational relationship, which remains grounded in trust, responsibility, and meaningful human mediation [5], [6], [7]. Recent analyses of large language models in education have emphasized that their impact depends less on technical capability than on pedagogical framing and institutional responsibility [3], [8]. This tension raises a central pedagogical question: under what conditions can AI-enabled interaction systems function as instruments of educational accompaniment without replacing, weakening, or distorting the human relationship at the core of education?

This paper adopts an explicitly human-in-command stance, understood here as a pedagogical interpretation of human oversight and accountability in educational AI governance [9], [10], according to which responsibility for educational processes remains with adults and educational communities, while AI operates as a constrained and intentional mediation. From this perspective, AI is not conceived as an autonomous empathic subject, nor as a surrogate for human care, but as an enabling infrastructure designed to support orientation, reflection, and transition toward authentic human interaction. Such a position is particularly relevant in contexts where ethical concerns and governance requirements are increasingly central to discussions on AI in education [3], [11]. In particular, the risk of unintended substitution dynamics and over-reliance has been raised as a concern in guidance on generative AI for education [3].

Within this framework, we introduce the concept of AI-enabled pedagogical accompaniment, understood as a structured process that supports learners in clarifying goals, articulating questions, and identifying next steps, while consistently orienting interaction toward teachers, peers, and communities of practice. The conceptual core of this approach is the notion of a relational bridge: a sequence of micro-mediations that lowers the threshold of access to the educational relationship and facilitates transitions from mediated interaction to direct human engagement. Rather than fostering prolonged or exclusive interaction with the AI system, the relational bridge is explicitly designed to be temporary, directional, and bounded.

A critical gap in current research concerns the operationalization and evaluation of such approaches. Many studies on AI in education focus either on system performance or on high-level ethical principles [1], [3], [4], [12], while fewer contributions provide replicable design principles linked to observable indicators that can be applied and assessed in real educational settings. Without such indicators, it becomes difficult for educators, institutions, and policymakers to distinguish between supportive mediation and substitutive dynamics, or to identify early signals of misuse or unintended consequences [9], [10].

To address this gap, this paper proposes a replicable framework of design principles and observable indicators for AI-enabled pedagogical accompaniment. The framework is grounded in the Amico project, an AI-mediated interaction system developed as a digital agent and, where feasible, as a social robot [13], [14], explicitly constrained by pedagogical and ethical requirements. The system was designed to support non-clinical educational contexts, particularly in technical and vocational education, where learners often face challenges related to self-efficacy, orientation, and continuity of engagement.

Accordingly, this paper should be read primarily as a design and accountability contribution: it presents a prototype-informed framework, a principle-to-indicator mapping, and an initial protocol for evidence generation. Its purpose is to enable replicable validation in future doctoral research and through international collaboration, rather than to report definitive outcomes.

A distinctive feature of the system is its dual-mode configuration: AmicoMio, oriented toward technical clarity, structured guidance, and task-focused support; and AmicoTuo, oriented toward reflective dialogue, supportive questioning, and meaning-making. These two modes are not conceived as competing approaches, but as complementary configurations addressing different educational needs and interaction moments. Their comparison provides a concrete basis for examining how interaction style influences acceptance, perceived usefulness, and transition toward human mediation across cultural contexts [14].

The study adopts an exploratory, mixed-methods approach and reports preliminary comparative evidence from pilot activities conducted in Italy and China. While the findings are non-conclusive, they enable the identification of design principles, corresponding observable indicators, and risk thresholds that can inform responsible deployment and future validation at scale. By translating abstract ethical and pedagogical commitments into measurable and actionable criteria, the paper aims to contribute both to research on AI in education and to the practical work of educators and institutions navigating the integration of AI-enabled systems.

This paper makes three contributions. (1) It defines AI-enabled pedagogical accompaniment operationalized as a relational bridge—a directional interaction device designed to scaffold learners toward human contact rather than substitute it. (2) It provides a compact, replicable mapping from design principles to observable indicators (log-, rubric-, survey-, and observation-based) to enable pedagogical accountability in real settings. (3) It reports design-based evidence from the Amico dual-mode prototype and frames Italy–China as a cross-context stress test for bounded, governance-aware educational mediation.

## II. Conceptual Framework

The conceptual framework of this study is grounded in a pedagogical interpretation of AI as a mediated educational infrastructure, rather than as an autonomous instructional or relational agent. This position emerges from the convergence of three foundational assumptions: (1) education is intrinsically relational and cannot be reduced to information delivery or performance optimization; (2) responsibility for educational processes remains with adults and educational communities; and (3) AI-mediated interaction systems can contribute meaningfully to education only if they are explicitly constrained by pedagogical, ethical, and governance requirements.

### A. *AI-Enabled Pedagogical Accompaniment*

Within this framework, AI-enabled pedagogical accompaniment is defined as a structured process that supports learners in orientation, reflection, and decision-making while consistently directing interaction toward meaningful human relationships. Accompaniment is understood not as continuous support or affective substitution, but as ex-ducere: guiding individuals toward the discovery and development of their capacities [5], [15], [16], [17] within a network of relationships that includes teachers, peers, and communities of practice [7].

This interpretation explicitly rejects both technocentric and substitutionist models [3], [4], [9], in which AI is framed as an empathic surrogate or as an autonomous tutor. Instead, AI is conceived as a temporary and directional mediation, designed to lower barriers to participation, clarify goals, and facilitate transitions toward human interaction. The educational value of such systems lies not in prolonged engagement with the AI itself, but in their capacity to activate and reorient human educational dynamics [5], [6], [7]. This conception of accompaniment aligns with pedagogical approaches that frame technology as an extension of human responsibility and meaning-making rather than as an autonomous substitute [18].

### B. *Human-in-Command and Adult Responsibility*

A central pillar of the framework is the principle of human-in-command [9], [10], according to which adults—teachers, tutors, and educational institutions—retain responsibility for educational decisions, relational boundaries, and ethical oversight. In this perspective, AI systems operate under predefined constraints that make their limits explicit, prevent ambiguity regarding their status, and avoid anthropomorphic misinterpretations [3], [4].

Human-in-command implies that AI-mediated interactions must be designed to:

- maintain transparency regarding system identity and capabilities;
- support, rather than replace, adult judgment;

- include explicit thresholds for escalation or hand-off to human actors;
- prevent exclusive or dependency-oriented use patterns [3], [9].

This stance is particularly relevant in educational contexts where learners may experience vulnerability related to self-efficacy, motivation, or belonging [6], [16], [17]. By embedding adult responsibility into system design, the framework seeks to ensure that AI acts as a supportive mediation [5], [7], not as an alternative relational pole.

### C. The Relational Bridge as Educational Mediation

The conceptual core of the framework is the notion of a relational bridge, defined as a sequence of micro-mediations that reduce the threshold of access to the educational relationship and facilitate transitions toward direct human engagement. The relational bridge is not a metaphorical construct, but an operational concept that informs system behavior, interaction design, and evaluation criteria.

A relational bridge is characterized by three properties:

1) Temporality: the AI-mediated interaction is explicitly limited in scope and duration;
2) Directionality: interactions are oriented toward concrete next steps involving human actors (teachers, peers, communities);
3) Boundedness: system behavior is constrained by pedagogical purposes and ethical safeguards.

In practical terms, the relational bridge supports learners in articulating questions, clarifying intentions, and identifying actionable steps, while progressively encouraging contact with relevant human resources. The effectiveness of the bridge is therefore assessed not by engagement with the AI system itself, but by observable transitions toward educational participation, agency, and responsibility.

### D. From Principles to Observable Indicators

A critical limitation of many discussions on AI in education lies in the gap between normative principles and empirical evaluation [3], [9], [10]. Ethical guidelines and pedagogical values are often articulated at a high level of abstraction, making it difficult for educators and institutions to assess whether a given system supports or undermines educational aims in practice.

To address this gap, the present framework emphasizes the translation of pedagogical principles into observable indicators. Core design principles—such as transparency, scaffolding toward human relationships, prevention of dependency dynamics, and data minimization—can be translated into indicators that are observable, measurable, or qualitatively assessable in real educational settings [1], [3], [12]. In this study, the principle-to-indicator mapping is treated as an accountability device: it makes the intended pedagogical function of the system testable through concrete traces (interaction logs), structured judgments (rubrics), and participant perceptions (short measures and feedback) [3], [9]. Observable indicators are treated as signals supporting professional judgment, not as exhaustive KPIs. For this reason, the framework deliberately combines light quantitative traces with qualitative rubrics and facilitator observation, in order to avoid reducing pedagogical accompaniment to what is easiest to measure.

Table I summarizes a compact operational mapping used to guide both design and analysis.

TABLE I. PRINCIPLE–INDICATOR–OBSERVATION MAPPING [a]

| Principle (design intent) | Observable indicator (example) | How it is observed (data source)[b] |
|---|---|---|
| Transparency of identity and limits | Misinterpretation rate (anthropomorphic framing; “AI as person”) | **Survey/Obs** (post-task items + facilitator notes) |
| Scaffolding toward humans (“relational bridge”) | Transition-to-human prompts and follow-through signals | **Log/Obs** (prompt occurrences + reported or facilitated next-step actions) |
| Directional micro-mediations | Micro-action formulation / completion rate | **Log/Rubric** (presence of actionable steps; rubric on specificity) |
| Prevention of dependency dynamics | Looping / extension attempts; exclusive-use signals | **Log** (repeated sessions; ignoring closure prompts) |
| Communicative autonomy support | Goal/question articulation quality (pre/post) | **Rubric** (clarity, specificity, agency markers) |
| Data minimization and purpose limitation | Sensitive-data solicitation frequency (should be near zero) | **Log** (content flags; redirection triggers) |

a. Compact operational mapping (principle → indicator → observation method) used to guide both design and analysis.

b. Observation codes: Log = interaction logs; Rubric = structured rating of transcripts; Survey = short self-report items; Obs = facilitator observation notes.

TABLE II. RELATIONAL BRIDGE LOOP

| **Relational Bridge Loop** | |
|---|---|
| ***Step*** | ***Description*** |
| 1 | Entry. learner expresses confusion, goal, situation |
| 2 | Micro-mediation. clarify goal: generate 1 to 3 micro-actions |
| 3 | Directionality. identify human target (teacher, peer, community) |
| 4 | Exit. closure prompt + next step plan |
| 5 | Accountability. Log, rubric, survey, indicators, risk check |
| 6 | If risk signal. reduce interaction + recommend human contact |

### E. Cross-Cultural Applicability

Finally, the framework is intentionally designed to support cross-cultural analysis. While the core principles of human responsibility and relational mediation are shared, their implementation and interpretation may vary across educational systems and cultural contexts, including differences shaped by educational reforms and pedagogical traditions in China [19], [20]. The comparative dimension of the study (Italy–China) is therefore not aimed at establishing uniform outcomes, but at examining how contextual factors influence acceptance, usage patterns, and perceived risks of AI-enabled pedagogical accompaniment [19], [20].

By foregrounding principles and indicators rather than fixed behaviors, the framework seeks to remain adaptable to diverse institutional settings while preserving a coherent pedagogical orientation [5], [7].

## III. System Design and Dual-Mode Prototype (Amico)

### A. *Design Rationale and Scope*

The Amico system was conceived as an AI-mediated interaction device explicitly designed for non-clinical educational contexts [3], with particular attention to technical and vocational education settings. The design rationale does not aim at creating an autonomous tutor or a relational surrogate [3], but at implementing a pedagogically constrained mediation that supports orientation, reflection, and transition toward human educational relationships.

From the outset, the system was developed according to a set of non-negotiable constraints derived from the conceptual framework: human-in-command governance, transparency of system identity and limits, prevention of dependency dynamics [9], [10], data minimization [21], and explicit orientation toward teachers, peers, and communities of practice. These constraints informed all design decisions, including interaction style, response structure, and termination or escalation rules [3], [9].

The scope of Amico is intentionally limited. It does not provide curricular instruction, psychological counseling, or affective companionship [3], [9]. Amico is not an autonomous tutor and does not optimize for content mastery through personalization. Its scope is pedagogical micro-mediation under adult responsibility, focusing on (i) supporting actionable next steps, (ii) lowering barriers to participation, and (iii) facilitating transitions toward teachers, peers, or communities of practice. Accordingly, success criteria prioritize relational and governance outcomes over learning gains. Its function is to support learners in clarifying questions, identifying next steps, and preparing meaningful engagement with human actors. In this sense, the system is best described as an educational companion [2], [12], where "companion" refers to accompaniment toward action and relationship, not to continuous presence or emotional substitution.

### B. *Dual-Mode Interaction as a Pedagogical Choice*

A distinctive feature of the Amico system is its dual-mode configuration, implemented through two complementary interaction modes: AmicoMio and AmicoTuo. This duality is not a technical artifact, but a deliberate pedagogical choice intended to address different educational needs and interaction moments [5], [7].

AmicoMio is oriented toward technical clarity and structured guidance. Its interaction style emphasizes precision, step-by-step reasoning, explicit framing of tasks, and concrete suggestions for action. This mode is particularly suited to moments in which learners seek orientation, procedural understanding, or validation of their approach to a problem.

AmicoTuo is oriented toward reflective and supportive interaction. Its responses prioritize maieutic questioning, reformulation, and meaning-making, encouraging learners to articulate intentions, values, and goals. This mode is designed to support moments of uncertainty, motivational fatigue, or decision-making, without adopting a therapeutic or affective stance.

The two modes are not presented as mutually exclusive alternatives, nor as "better" or "worse" solutions. Instead, they represent complementary configurations within the same pedagogical framework. The availability of both modes enables exploration of how interaction style influences acceptance, perceived usefulness [4], [12], [22], and transitions toward human mediation across different cultural and educational contexts [19], [20].

### C. *Operationalizing the Relational Bridge*

The concept of the relational bridge, introduced in the conceptual framework, is operationalized in Amico through specific interaction patterns and constraints. Each interaction is designed to function as a micro-mediation that lowers the threshold of access to the educational relationship and supports movement toward concrete human engagement [7].

In practice, this operationalization involves:

- Explicit prompts that encourage identification of relevant human resources (teachers, tutors, peers);
- Formulation of micro-actions that can be undertaken outside the AI interaction (e.g., preparing a question, contacting a peer, scheduling a meeting);
- Conversational closures that emphasize transition rather than continuation of AI-mediated dialogue.

A learner reports post-lesson confusion ("I think I understood, but I cannot explain it; I'm afraid to ask"). The system first performs a micro-mediation by requesting a concrete goal ("What do you want to be able to do in one sentence?") and proposing 1–3 micro-actions (e.g., "write a two-line question," "identify the step where uncertainty begins," "prepare a diagram/photo of the point of doubt"). It then enforces directionality by asking the learner to select a human target (teacher/peer/community) and generates a short message the learner can reuse. The interaction ends with a closure prompt and an explicit next-step plan. A logged "transition attempt" is recorded when the learner selects a human target; a "follow-through" signal is recorded when the learner reports (or a facilitator confirms) that the step was executed.

Crucially, the system avoids open-ended conversational loops [3], [9]. Interaction sequences are intentionally bounded and oriented toward exit points that redirect learners to human contexts. The success of the relational bridge is therefore not evaluated through prolonged AI engagement, but through observable transitions toward educational participation and responsibility [5], [16], [17].

### D. *Transparency, Boundaries, and Risk Prevention*

Transparency regarding system identity and limits is a core design requirement. Amico consistently presents itself as an AI-based system with predefined capabilities and constraints, avoiding anthropomorphic language or representations that could foster misinterpretation [3], [4]. This transparency is reinforced through interaction cues that clarify the system's role as a support tool rather than an authoritative or empathic agent.

To prevent dependency dynamics [3], [9], the system incorporates usage boundaries and risk signals. These include limitations on interaction frequency and duration [3], detection of repetitive or exclusive use patterns, and prompts that explicitly suggest human contact when predefined thresholds are reached. Such mechanisms are not conceived as control features, but as pedagogical safeguards aligned with adult responsibility and institutional oversight.

Data handling follows a principle of minimization and purpose limitation [21], [23]. Only information strictly necessary for the educational interaction is processed, and sensitive content is handled with caution, triggering redirection to human support when appropriate [3], [9]. These measures are essential to maintaining trust and ensuring that the system remains compatible with diverse governance and regulatory environments.

### E. *Design Implications for Evaluation*

The design of the Amico system is tightly coupled with its evaluation strategy. Each design choice — particularly the dual-mode configuration and the operationalization of the relational bridge — is explicitly linked to observable indicators that can be assessed in real educational settings [1], [12]. This alignment enables systematic examination of how different interaction modes and constraints influence learner behavior, acceptance, and transition toward human mediation [22].

By foregrounding design decisions that are pedagogically motivated and empirically observable, the system provides a concrete basis for design-based research and iterative refinement. The following section describes the methodological approach adopted to explore these dynamics and to generate preliminary evidence across different cultural contexts [19], [20].

### F. *Age-Sensitive Configuration and Model Adaptation*

Beyond interaction style, the Amico dual-mode prototype was configured to respond differently by age band, reflecting developmental differences in language, agency, and support needs. Age bands were aligned with the pilot's heterogeneous participant profile (children and adolescents, young adults, and adults). This age-sensitive design intent was implemented through (i) mode-specific instruction sets for AmicoMio and AmicoTuo and (ii) specialized datasets used for targeted adaptation of the supportive mode.

In particular, the supportive mode (AmicoTuo) was refined via iterative dataset generation and fine-tuning cycles to better align responses with congruent communication principles and the Edulife Value Cycle, while maintaining the human-in-command safeguards described above. In parallel, the assistant-level sampling parameters (temperature and top-p) were tuned per mode to balance determinism and linguistic variety: lower-variance settings were preferred for task-oriented guidance (AmicoMio), whereas moderately higher-variance settings were used for reflective prompts (AmicoTuo) when they did not compromise clarity or boundaries. These parameters were treated as part of the system design surface and therefore as factors to be controlled and reported in subsequent validation studies. [24]

## IV. Methods

### A. *Research Design*

The study adopts an exploratory, mixed-methods research approach [1], [12], [22]. This choice reflects the dual objective of the research: (a) to investigate the feasibility and acceptability of an AI-enabled pedagogical accompaniment system under real educational constraints, and (b) to derive replicable design principles and observable indicators to inform future validation and scaling.

An exploratory mixed-methods approach is particularly suited to early-stage educational innovation, as it enables iterative refinement while maintaining explicit links between theory, design decisions, and empirical observation. Quantitative and qualitative methods were combined to capture both measurable interaction patterns and participants' interpretations, meanings, and perceived risks.

### B. *Contexts and Participants*

Pilot activities were conducted in Italy and China, selected to reflect diversity in institutional culture while maintaining comparability at the level of educational purpose [11], [19], [20]. In Italy, the exploratory sessions were conducted in collaboration with Edulife S.p.A. Impresa Sociale and Fondazione Edulife; in China, they were conducted in collaboration with Zhejiang Yizhong Intelligent Technology Co., Ltd. The pilot involved N = 30 participants (N_IT = 12; N_CN = 18) and a total of 20 bounded sessions (10 in Italy; 10 in China). The primary focus was on technical and vocational education settings, where learners often encounter challenges related to orientation, self-efficacy, and continuity of engagement [6], [16], [17].

Participants were intentionally heterogeneous by age and role, spanning three broad age groups (6–18 years, 20–40 years, and 40–65 years) and including learners as well as adult educational actors (e.g., teachers and parents) engaged in structured learning or support contexts. Participation was voluntary, and all activities were conducted in non-clinical contexts [3], [9]. The study did not target individuals with diagnosed psychological conditions, and the system was not presented as a therapeutic or counseling tool.

Given the exploratory nature of the study, the sample size was not intended to support statistical generalization. Instead, it enabled the identification of patterns, contrasts, and critical episodes relevant to system design and pedagogical mediation.

### C. *Procedure and Interaction Scenarios*

Participants interacted with the Amico system in structured scenarios designed to reflect realistic educational use cases [3]. These scenarios included moments of task clarification, decision-making, reflection on next steps, and preparation for interaction with teachers or peers.

Each participant engaged with both interaction modes—AmicoMio and AmicoTuo—either sequentially or in parallel sessions. The order of exposure was varied to reduce sequence effects. Interaction sessions were intentionally bounded in duration, in line with the system's design constraints [3], [9], and were followed by brief reflective activities.

Throughout the pilot, educators or facilitators maintained an oversight role, consistent with the human-in-command principle [9], [10]. Participants were explicitly informed of the system's identity, purpose, and limits prior to interaction [9].

### *D. Data Collection*

Data collection focused on linking design principles to observable indicators across multiple sources [1], [12]:

- Interaction logs: anonymized records capturing interaction frequency, duration, and mode usage.
- Structured rubrics: qualitative assessments of communicative autonomy, clarity of goals, and articulation of next steps, applied to selected interaction transcripts.
- Pre/post self-report measures: short items assessing perceived engagement, self-efficacy, and proxy belonging indicators (short self-report items) [16], [17].
- Post-interaction questionnaires: participant perceptions of usefulness, clarity, and appropriateness of each interaction mode.
- Qualitative feedback: brief open-ended responses and facilitator observations focused on acceptance, misunderstandings, and perceived risks.

Given the pilot scope, facilitator notes and brief post-task items are treated as indicative and will require confirmation with larger samples and validated instruments.

In addition, risk signals were monitored, including repetitive or exclusive use patterns and reluctance to transition toward human interaction [3], [9], [10].

### *E. Indicators and Analytical Strategy*

Analysis was guided by a predefined mapping between design principles and observable indicators [1], [12]. Quantitative data were analyzed descriptively to identify trends and contrasts between interaction modes and contexts. Given the sample size, inferential statistics were not emphasized; where appropriate, simple pre/post comparisons were used to explore directional changes.

Qualitative data were analyzed through thematic analysis, focusing on patterns related to orientation, agency, transition toward human mediation, and perceived boundaries of the system. Attention was paid to culturally influenced interpretations and to episodes where system constraints prompted redirection to human actors.

The integration of quantitative and qualitative findings enabled triangulation and supported the identification of conditions of effectiveness, rather than claims of efficacy.

### *F. Ethical Considerations*

The study was conducted in accordance with principles of non-clinical educational research and ethical responsibility [3], [9]. Participants were informed about the nature of the system, the scope of data collection, and their right to withdraw at any time. Data handling followed principles of minimization and purpose limitation, and no sensitive personal data were intentionally solicited [21], [23].

Author positionality: The author holds leadership roles in the partner organizations (co-founder of Edulife and Deputy General Manager at Zhejiang Yizhong Intelligent Technology Co., Ltd.). To mitigate potential bias, the pilot relied on facilitator oversight consistent with the human-in-command principle, triangulation across logs, rubrics, surveys, and open-ended feedback, and the analysis is reported as exploratory design-based evidence rather than as an efficacy evaluation.

Explicit safeguards were implemented to prevent relational dependency [3], [9], including bounded interaction sessions and prompts encouraging human contact. The research design prioritized transparency, adult responsibility, and institutional oversight throughout all phases of the pilot [9], [10].

## V. Initial Observations (Design-Based Insights)

These observations are reported to inform design iteration and indicator feasibility; they are not intended as efficacy evidence.

### *A. Overview of Observed Patterns*

Results are presented as preliminary and exploratory, focusing on patterns emerging from the comparison between interaction modes (AmicoMio and AmicoTuo) and across cultural contexts (Italy and China). Given the pilot nature of the study and the limited sample size, findings are reported descriptively and interpretatively, without claims of statistical generalization.

Overall, participants were able to engage with the system as intended, recognizing its role as a supportive and bounded educational mediation rather than as an autonomous tutor or relational substitute [3], [9]. Across contexts, no major misunderstandings regarding system identity were observed during the pilot sessions, and transparency mechanisms appeared effective in preventing anthropomorphic interpretations [3], [4], [10]. These observations are based on facilitator notes and brief post-task survey items and should be confirmed with larger samples and validated instruments.

### *B. Acceptance and Perceived Usefulness*

Both interaction modes were generally perceived as useful, though for different educational purposes.

- AmicoMio was more frequently associated with clarity, structure, and task-oriented usefulness. Participants reported that this mode helped them organize thoughts, understand procedural steps, and prepare concrete questions for teachers or peers.
- AmicoTuo was more frequently associated with reflective value, meaning clarification, and motivational support. Participants highlighted its usefulness in moments of uncertainty, indecision, or reduced confidence.

Across the full sample, participants tended to perceive the two modes as complementary rather than alternative, indicating that their usefulness depended on the interaction moment rather than on personal preference alone [3], [12].

### C. Observable Indicators of Educational Mediation

Analysis of observable indicators revealed several consistent patterns:

- Transition-to-human contact: Multiple interaction sequences included explicit prompts directing participants toward teachers, peers, or relevant human actors. Qualitative debriefing and facilitator observations indicate that such transitions occurred in several cases; however, the pilot did not implement a formalized mechanism to systematically verify human follow-through. For this reason, transition-to-human follow-through is not reported here as a quantified metric. The instrumentation and validation of this indicator constitute a central objective of the subsequent doctoral research phase.

- Micro-action completion: Participants frequently identified concrete next steps (micro-actions), such as drafting questions, revising plans, or scheduling follow-up activities. Completion rates were higher when interactions included structured prompts aligned with task clarity.

- Communicative autonomy: Qualitative rubric analysis indicated improvements in the articulation of goals and questions between pre- and post-interaction phases, particularly following AmicoTuo sessions, where reflective questioning was emphasized.

These indicators suggest that the system functioned as a directional mediation, supporting orientation and action rather than prolonged AI engagement [3], [16], [17].

### D. Usage Patterns and Boundary Effects

Usage data confirmed that interaction sessions remained bounded in duration and frequency, consistent with design constraints. No clear patterns of excessive or exclusive use emerged during the observed sessions.

Risk signals related to dependency—such as repeated attempts to extend interaction beyond suggested closure points—were rare and, when present, were effectively addressed through system prompts encouraging human contact. This suggests that the implemented safeguards contributed to maintaining the system within its intended pedagogical scope [3], [9], [10].

### E. Cross-Cultural Observations

While core patterns were consistent across contexts, contextual differences emerged in the preferred interaction mode:

- In the Italian context, participants more frequently emphasized the reflective and dialogical aspects of AmicoTuo, valuing opportunities for meaning-making and personal articulation.

- In the Chinese context, participants more frequently emphasized the relational harmony and contextual sensitivity of AmicoTuo, valuing its supportive and reflective stance in educational and family-related scenarios. AmicoMio was nonetheless appreciated for its technical reliability in work-oriented contexts.

These differences did not indicate divergent outcomes, but rather context-sensitive preferences in how pedagogical mediation is experienced and valued [11], [19], [20]. Importantly, in both contexts, participants recognized the system's orientation toward human mediation and did not report expectations of substitution or emotional dependency [3], [9].

### F. Summary of Preliminary Findings

Taken together, the results indicate that the Amico system:

- was accepted as a legitimate educational support across contexts;

- supported observable transitions toward human interaction;

- enabled complementary use of technical and reflective interaction modes;

- operated within intended boundaries without fostering dependency dynamics.

These findings provide initial empirical support for the feasibility of AI-enabled pedagogical accompaniment when grounded in explicit design principles and evaluated through observable indicators [1], [3], [9].

## VI. Discussion

The present study set out to explore the conditions under which AI-enabled interaction systems can function as pedagogical accompaniment devices rather than as substitutive instructional or relational agents. The findings, while preliminary, offer several insights that contribute to current debates on AI in education, particularly regarding design accountability, human responsibility, and cross-cultural applicability [1], [3], [12].

### A. Usage Patterns and Boundary Effects

A first key implication concerns the role of pedagogical mediation. The results suggest that the perceived usefulness of the Amico system did not primarily depend on technological novelty, but on the clarity of its pedagogical orientation and boundaries [1], [12]. Participants across contexts recognized the system as a supportive mediation, oriented toward action and human engagement, rather than as an autonomous authority or companion [3], [9].

This finding reinforces the argument that, in educational contexts, the value of AI-enabled systems lies less in their capacity to simulate human traits and more in their ability to structure reflective and action-oriented processes. By explicitly avoiding open-ended interaction and by embedding exit points toward human actors, the system aligns with a conception of education as relational and responsibility-based [3], [4], [10].

### B. Dual-Mode Interaction as a Design Asset

The dual-mode configuration (AmicoMio and AmicoTuo) emerged as a particularly relevant design feature. Rather than generating confusion or fragmentation, the availability of two complementary interaction styles supported differentiated educational needs. Technical clarity and reflective support were not experienced as competing functions, but as context-dependent resources [12].

This suggests that interaction style is a meaningful design dimension in AI-enabled educational systems and should be treated as such in both design and evaluation. The results indicate that offering multiple, pedagogically grounded interaction modes can enhance adaptability without sacrificing coherence, provided that all modes remain aligned with the same ethical and educational constraints [3], [9], [10].

### *C. Observable Indicators and Pedagogical Accountability*

A central contribution of this work lies in the explicit linkage between design principles and observable indicators. The use of indicators such as transition-to-human contact, micro-action completion, and communicative autonomy enabled a form of evaluation that goes beyond system performance metrics and addresses educational purpose [1], [12].

This approach supports a notion of pedagogical accountability, whereby AI-enabled systems can be assessed in terms of how they shape participation, agency, and responsibility [6], [16], [17]. Importantly, such indicators also make it possible to identify early signals of misalignment or risk, enabling timely intervention and iterative refinement [3], [9], [10].

For policymakers and educational institutions, this linkage offers a practical pathway to operationalize high-level ethical commitments into evaluable criteria [9], [10], [21], [23], [25].

### *D. Cross-Cultural Interpretation and Context Sensitivity*

The comparative dimension of the study highlights the importance of context sensitivity. While core principles were shared across Italy and China, differences in preferred interaction modes reflected broader educational cultures and expectations. These differences should not be interpreted as barriers to transferability, but as indicators of the need for adaptive implementation [19], [20].

By focusing on principles and indicators rather than fixed behaviors, the framework supports contextual adaptation while maintaining a coherent pedagogical stance. This is particularly relevant in international and cross-cultural deployments of AI-enabled educational systems, where uniform solutions are unlikely to be effective or appropriate [3], [9].

### *E. Implications for Early-Stage Research and Practice*

From a methodological perspective, the study illustrates the value of exploratory research in the early stages of AI-enabled educational innovation. Rather than aiming for definitive validation, such approaches enable the identification of conditions of feasibility, acceptance, and risk, which are essential prerequisites for responsible scaling [3], [9], [10].

For practitioners, the findings suggest that AI-enabled pedagogical accompaniment can be meaningfully integrated into educational practice when it is:

- explicitly bounded and directional;
- transparently positioned as a support tool;
- evaluated through indicators aligned with educational aims [3], [9], [10].

### *F. A Collaborative Research Agenda (Call for Action)*

To advance the proposed framework beyond conceptual articulation and exploratory prototyping, a coordinated and methodologically rigorous research agenda is required. The next phase of development should prioritize structured collaboration, cross-cultural validation, and transparent governance mechanisms. In particular, future work should include:

- The implementation of a shared data collection protocol integrating interaction logs, structured rubrics, and survey instruments, in order to ensure comparability and methodological consistency across research sites.
- The cross-cultural validation of design principles and observable indicators through partnerships involving institutions in Italy, China, and additional international contexts.
- The development of a governance and ethics compliance package aligned with both European Union and Chinese regulatory frameworks, ensuring responsible deployment and institutional accountability.
- The public release of replicable research materials, including validated rubrics, item banks, and structured observation codes, to facilitate cumulative knowledge building and independent replication.

## VII. Limitations and Future Work

### *A. Limitations*

This study presents several limitations that should be explicitly acknowledged. First, the research is based on an exploratory pilot with a limited sample size and heterogeneous educational contexts. As such, the findings are not intended to support statistical generalization or causal claims. Rather, they provide indicative patterns relevant to design feasibility, acceptance, and pedagogical mediation.

Second, part of the evidence relies on self-reported measures and qualitative interpretations, which may be influenced by contextual factors, participant expectations, or facilitator presence. Some observations rely on facilitator notes and brief post-task items and should be confirmed with larger samples and validated instruments. While triangulation across data sources was employed to mitigate these effects, further research with more robust measurement instruments is required.

Third, the comparative dimension (Italy–China), while conceptually central, is based on initial observations rather than controlled cross-cultural experimentation. Differences in institutional settings, educational traditions, and implementation conditions limit direct comparability and call for cautious interpretation.

Finally, the current study focuses on short-term interaction patterns and immediate transitions toward human mediation. Longitudinal effects, including sustained impact on learner agency, educational participation, or potential unintended consequences, were beyond the scope of this pilot.

### B. *Directions for Future Research*

Future work will aim to extend and consolidate the present findings along several complementary directions.

First, the framework of design principles and observable indicators will be subjected to larger-scale validation across diversified educational contexts, with particular attention to technical and vocational education settings. This will include more systematic sampling strategies and the use of quasi-experimental designs where appropriate.

Second, future studies will explore longitudinal dynamics, examining how AI-enabled pedagogical accompaniment influences learner trajectories over time, including transitions between mediated interaction and sustained human engagement. Such studies are essential to assess both effectiveness and risk prevention [3], [12], [23].

Third, the dual-mode configuration will be further refined and dynamically adapted, investigating conditions under which different interaction styles are most appropriate and how mode switching can be guided by pedagogical intent rather than user preference alone [12], [21], [23].

Fourth, the comparative dimension will be expanded through collaborative research networks involving educators and institutions in different cultural and regulatory environments. This collaborative approach is intended to support shared validation, contextual adaptation, and responsible knowledge transfer [19], [20].

Finally, future work will deepen the integration of ethical governance and institutional oversight mechanisms, translating high-level principles into operational protocols that can be adopted by schools, training centers, and educational authorities [3], [9], [10], [21], [23], [25].

## VIII. Conclusion

This paper argues that the educational value of AI-enabled interaction systems does not primarily depend on sophistication or "human-likeness," but on accountable pedagogical mediation: transparent limits, bounded interaction, and an explicit orientation toward human relationships. Within a human-in-command stance, we introduced AI-enabled pedagogical accompaniment as a directional, temporary, and safeguarded mediation, operationalized through the concept of a relational bridge.

The main contribution is a replicable framework that links design intent to observable indicators—transition-to-human signals, micro-action formulation, communicative autonomy, and early risk markers—so that educators and institutions can evaluate whether an AI system supports education or drifts into substitution and dependency dynamics. The Amico dual-mode prototype is presented as a concrete instantiation of this approach, designed to be useful in technical and vocational contexts while remaining ethically bounded and governance-compatible.

Importantly, this work does not claim efficacy. Instead, it provides a practical accountability toolkit and a research agenda for systematic validation in real educational environments. The next step is collaborative evidence generation: shared protocols, comparable indicators, and context-sensitive implementations across institutions and cultures. If adopted, this approach can help shift the debate from abstract promises and fears toward measurable, responsible, and human-centered educational practice.

The proposed next phase—implemented as a doctoral research program—will operationalize this agenda through shared measurement protocols, cross-context validation, and longitudinal evidence generation under human-in-command governance.